\documentclass[12pt,psfig]{iopart}
\usepackage{graphicx} 
\usepackage{pdfpages}
\usepackage{amsfonts}

\begin{document}
\title{Numerical solutions for phase noise due to pointing jitter with the LISA telescope}

\author{Jean-Yves Vinet, Nelson Christensen, Nicoleta Dinu-Jaeger, Michel Lintz, Nary Man and Mikha\"el Pichot} 
\address{ARTEMIS, Universit\'e C\^ote d'Azur, Observatoire de la C\^ote d'Azur, C.N.R.S., Nice 06304 France}
%\index[aindx]{Author, F.} % or \aindx{Author, F.}
%\index[aindx]{Author, S.} % or \aindx{Author, S.}

%\maketitle
%
\begin{abstract}
The aim of the Laser Interferometer Space Antenna (LISA) is to detect gravitational waves through a phase modulation in long (2.5 Mkm) laser light links between spacecraft. Among other noise sources to be addressed are the phase fluctuations caused by a possible angular jitter of the emitted beam. The present paper follows our preceding one (Vinet J-Y {\it et al} 2019 {\it Class. Quant. Grav.} {\bf 36} 205003)  based on an analytical study of the far field phase.  We address here a numerical treatment of the phase, to first order in the emitted wavefront aberrations, but without any assumptions on the static bias term. We verify that, in the phase change, the higher order terms in the static mispointing are consistent with the results found in our preceding paper.

\end{abstract}
%==============================================================
\section{Introduction}
After the successful observations by the ground based gravitational wave (GW) detectors operated by the LIGO Scientific Collaboration and the Virgo Collaboration in US and in Europe~\cite{LVC}, a new impulse has been given to an old and ambitious project, supported successively by the NASA and the ESA, called LISA (Laser Interferometer Space Antenna)~\cite{LISA}. LISA aims at detecting GWs in the very low frequency band (milliHertz), where a number of signals are 
%expected from all events involving directly or indirectly supermassive  
expected, particularly from events involving directly or indirectly supermassive black holes.
Continuous signals emitted by binary compact stars or pulsars are also in that low frequency band. 

The principle of the LISA Mission~\cite{CRAS} is to read the GW signal in the propagation time (or equivalently phase) of a light beam from an emitter laser to a receiver photodetector, both in heliocentric orbits, through a 2.5 Mkm path in space.  It is clear that addressing a target of meter size at such a distance is by itself a challenge. 
Among all possible spurious effects able to perturb the measurement, there is thus a possible misalignment of the emitted
beam due to any permanent (mechanical) or transient (intrinsic laser jitter) mispointing. We have addressed the question in a recent paper~\cite{CQGA}, where we used an analytical approach. We have shown that spurious effects are due to the conjunction of three elements: 
\begin{itemize}
\item
Some imperfections (aberrations) in the telescope used to enlarge the laser beam, which is necessary to temper the diffraction at a long distance. These aberrations are scaled by a length parameter $\sigma$ [m]
that expresses how much the emitted wavefront departs from an ideal plane;
\item
A static pointing error $\theta_0$;
\item
A jitter of the laser beam $\theta_1(t)$.
\end{itemize}
Analytical approaches are based on a first order theory in $\sigma/\lambda$, which seems reasonable, regarding the present state-of-the-art in mirror technology (stimulated, for example, by ground based GW detectors' demands). In our preceding approach, we furthermore adopted a first order treatment of $\theta_0$.  Higher orders can be addressed, but at the price an increasing complexity of the formulas, reducing their practical interest. 

In the present paper, we relax the first order approximation condition on $\theta_0$, which is possible using numerical methods. This numerical approach allows for the presentation of only some special examples, but provides a check on how the preceding (analytically obtained) numerical conclusions are relevant. This study should also be useful for any system where a laser beam is emitted and then detected at a very large distance.

The organization of this paper is as follows. In Sec.~\ref{sec:theory} we present the theoretical derivation for the phase error introduced by static and transient mispointing of the LISA telescope. Sec.~\ref{sec:numerical} contains the numerical calculation of the phase. A conclusion is given in Sec.~\ref{sec:conclusion}. Sec.~\ref{sec:annex} is an Annex that provides a comparison with the results presented in our previous publication~\cite{CQGA}.
%-----------------------------------------------------------------
\section{Theory}
\label{sec:theory}
We consider a special link from an emitter spacecraft (SC) to a receiver SC.
The emitter SC emits light of wavelength $\lambda$ through a telescope
of aperture (half-diameter) $a$.  The aim of this telescope is to increase the width of an initial
Gaussian beam up to the final Gaussian radius parameter $w$.  $(x,y,z)$ represents the coordinate system in which
$z$ is along the light propagation path; if $(x,y,0)$ are the coordinates
in the plane containing the emitting aperture, the emitted amplitude $A_0$ of the assumed Gaussian beam would be
(with $r\equiv \sqrt{x^2+y^2}$):
\begin{equation}
A_0(x,y,0) \ = \ \exp (-r^2/w^2) \ \  (r\le a), \  \  A_0(x,y,0)=0 \  \  (r>a)
\end{equation}
under ideal conditions.  Unfortunately, the mirrors building the telescope are not 
perfect and distortions of the phase surface, called aberrations, exist.  It has been shown~\cite{CQGA} that
those defects, coupled with mispointing and fluctuations (jitter) can cause a phase noise at reception.
It is essential to assess this noise and compare it to the sensitivity level required by LISA. 

We may represent the global wavefront aberration as a phase factor $k F(x,y)$ applied to $A_0$, in such a way that the actual emitted amplitude is now
\begin{equation}
\label{eq:amp_phase}
A(x,y,0) \ = \ A_0(x,y,0) \times \exp[i k F(x,y)] \ \  (k\equiv 2\pi/\lambda).
\end{equation}
It is conventional to expand aberrations in a circular pupil in a series of Zernike functions~\cite{BW}. 
The Zernike functions are :
\begin{eqnarray}
Z_n^{(m)}(\rho,\phi)\ = c_n^{(m)}\ R_n^{(m)}(\rho) \cos (m\phi), \\
Z_n^{(-m)}(\rho,\phi)\ = c_n^{(m)}\ R_n^{(m)}(\rho) \sin (m\phi),
\ \ (\rho\equiv r/a),
\end{eqnarray}
where the $R_n^{(m)}$ are the Zernike polynomials~\cite{BW}:
\begin{equation}
R_n^{(m)}(\rho)\ = \  
\sum_{s=0}^{(n-m)/2}(-1)^s \frac{(n-s)!}{s![(n+m)/2-s]! [(n-m)/2-s]! }\rho^{n-2s}
\end{equation}
and $$
c_n^{(m)}\equiv \sqrt{\frac{2(n+1)}{\pi(1+\delta_{m,0})}} ~ .
$$
The polar coordinates are defined by $(x=a \rho \cos\phi, y=a \rho \sin\phi)$.
An expansion of $F$ in a series of Zernike functions is thus:
\begin{equation}
F(x,y) \ = \  \sum_{n,m} c_n^{(m)} \left[ \sigma_{n,m} \cos m\phi \ + 
\sigma'_{n,m}\sin m\phi \right]R_n^{(m)}(r/a)
\end{equation}
The $\sigma_{n,m},\sigma'_{n,m}$ have the following definitions:
\begin{equation}
\sigma_{n,m}\ \equiv \int_\Delta R_n^{(m)}(\rho)F(x,y) \cos (m\phi )\rho\, d\rho \,d\phi ~,
%\  \ (x=a \rho \cos\phi, y=a \rho \sin\phi)
\end{equation}
and
\begin{equation}
\sigma'_{n,m}\ \equiv \int_\Delta R_n^{(m)}(\rho)F(x,y) \sin (m\phi )\rho\, d\rho \,d\phi ~.
\end{equation}
The $\sigma$'s have thus the dimension of a length.  $\Delta$ is the disk of radius $a$ in the plane $z=0$.
%Moreover, we consider a possible mixing of a static mispointing error defined by the  angles ($\theta_0,\psi_0$), and a dynamic jitter 
We now consider, as we did in our preceding paper~\cite{CQGA}, that the beam is emitted with a mispointing error that contains both a static value defined by the angles ($\theta_0,\psi_0$), and a dynamic jitter
defined by angles ($\theta_1(t),\psi_1(t)$).
This gives an additional phase factor in Eq.~\ref{eq:amp_phase}:
\begin{equation}
 \exp\left[
i k x(\theta_0 \cos\psi_0+\theta_1\cos\psi_1)+i k y (\theta_0 \sin\psi_0+\theta_1\sin\psi_1) 
\right] \  \ ,
\end{equation}
which finally with ($x\equiv r \cos\phi,\, y\equiv r \sin\phi$) leads to the aberrated and mispointed
amplitude:
\begin{equation}
A(x,y,0) \ = \ A_0(x,y,0)\times e^{ikF(x,y)} \times e^{ikr\theta \cos(\phi-\psi)} ~,
\end{equation}
where
\begin{equation}\label{deftheta}
\theta \equiv \sqrt{\theta_0^2+\theta_1^2+2\theta_0 \theta_1\, \cos\delta\psi}
\ \ , \psi\equiv \arctan{\left[
\frac{\theta_0\sin\psi_0+\theta_1\sin\psi_1}{\theta_0\cos\psi_0+\theta_1\cos\psi_1}
\right]} ~,
\end{equation}
with $\delta\psi\equiv \psi_0-\psi_1$.
Now, if we consider the far propagated field amplitude $B(x,y,L)$ at a distance $L$, it is 
well known that it amounts (Fraunhofer regime) to computing the Fourier transform of $A(x,y,0)$
namely:
\begin{equation}
B(x,y,L) \ = \ - \ \frac{i}{\lambda L} \ \exp\left[ i\pi \frac{r^2}{\lambda L}\right]
\int_{\mathbb R^2} e^{ipx'+iqy'}A(x',y',0) \,dx' \,dy'
\end{equation}
with ($p\equiv kx/L,\, q\equiv ky/L$).  
For a geometry like that for LISA where the beam exiting one telescope has a waist $\sim a$, and the beam observed by the receiving telescope has an aperature radius also of $a$, the condition for the amplitude of the electric field to be uniform to better than $x\%$ across this area is 
\begin{equation}
z>\frac{10^{4/3} \pi  a^2}{\lambda  x^{2/3}} ~ .
\end{equation}
At the very long distance for the LISA arms, $L \sim$ 2.5 Mkm, the
amplitude inside a disk of radius a ($\sim$ 15cm), is practically constant and given
simply by 
\begin{equation}
B(0,0,L) \ = \ - \ \frac{i}{\lambda L} \int_{\mathbb R^2} A(x,y,0) \,dx \,dy ~,
\end{equation}
or, after a change ($x,y) \rightarrow (r,\phi)$:
\begin{equation}
B(0,0,L) \ = \ - \ \frac{i}{\lambda L} \int_0^\infty r\,dr\, \int_0^{2\pi} d\phi \,A_0(r,\phi,0)
e^{ikF(r,\phi)}e^{ikr\theta \cos(\phi-\psi)} ~.
\end{equation}
If we assume very weak aberration amplitudes ($\sigma_{n,m},\sigma'_{n,m} \ll \lambda$), 
we can expand to first order the first exponential factor, and write:
%we can restrict the formula to the first order in $kF$ and write:
\begin{equation}
\hspace*{-2cm} B(0,0,L)\ = \ - \ \frac{i}{\lambda L} \int_0^\infty r\,dr\, \int_0^{2\pi} d\phi \,
e^{ikr\theta \cos(\phi-\psi)} \,A_0(r,\phi,0)
\left[1 \, +ik F(r,\phi) \right]
\end{equation}
which yields 
\begin{equation}\label{DNN}
\hspace*{-2cm} B(0,0,L)\ = \ - \ \frac{ik a^2}{\lambda L} \left\{ D   
+ \ 
ik \sum_{n,m} i^m c_n^{(m)}  \left[ \sigma_{n,m} \cos(m\psi) + \  \sigma'_{n,m}\sin(m\psi) \right]N_{n,m}
\right\}
\end{equation}
where
\begin{equation}\label{defbigd}
D(\Omega,v) \equiv  \int_0^1 J_0(\Omega \rho) e^{-v \rho^2}\rho\,d\rho
\end{equation}
with $\Omega \equiv ka\theta$, $v \equiv a^2/w^2$,  and
\begin{equation}\label{defbign}
N_{n,m} (\Omega,v) \ \equiv \ 
 \int_0^1  J_m(\Omega \rho) e^{-v \rho^2}R_n^{(m)}(\rho) \, \rho\,d\rho ~,
\end{equation}
the $J_m$ being the Bessel functions of the 1$^{st}$ kind.
The parameter $w$ (the telescope's aperture $a$ being fixed) has an optimum value
resulting from a compromise between diffraction losses (too small $w$) and clipping losses (too large $w$).
The optimum value (see~\cite{CQGA}) is $a/w$=1.12. However clipping losses result in scattered light issues
which are attenuated by taking a slightly suboptimal value $a/w$=1.5. This is why in this sequel article
we compare the results obtained with these two possible options.
From Eq.~\ref{DNN} we see that (to first order in $F/\lambda$) only even terms in $m$ will contribute to the phase
of $B(0,0,L)$, and thus both $m$ and $n$ are even.  
Eventually the spurious phase is:
\begin{equation}
\hspace*{-2cm} \delta \Phi \ = \  k\sum_{n,m} (-1)^m c_{2n}^{(2m)} \left[ \sigma_{2n,2m}\cos(2m\psi)
+ \sigma'_{2n,2m} \sin(2m\psi) \right] \frac{N_{2n,2m}(\Omega,v)}{D(\Omega, v)}
\end{equation}
%=================================================================
\section{Numerical}
\label{sec:numerical}
As described above, for the evaluation of the noisy phase we essentially have to compute:
\begin{equation}\label{defr}
G_{n,m}(\Omega,v) \ \equiv \ \frac{N_{2n,2m}(\Omega,v)}{D(\Omega, v)}
\end{equation}
Integrals of the kind (Eqs.~\ref{defbigd} and~\ref{defbign}) can be easily numerically integrated by Simpson's rule.
An excellent precision is reached with a sampling rate of 1000.
%we have confirmed that this gives a relative accuracy that is much better than $10^{-9}$.
We have checked that integrals similar to the ones in Eqs.~\ref{defbigd} and~\ref{defbign}, but for which we have an analytical expression, can be computed with a relative accuracy much better than $10^{-9}$.
We recall that
the parameter $\Omega$ is $\Omega=ka(\theta_0+ \theta_1\cos\delta\psi)=\Omega_0+ka \theta_1\cos\delta\psi$,
(see Eq.~\ref{deftheta} at first order in $\theta_1$)
where $\Omega_0 = k a \theta_0$ is due to the static mispointing (we use the value 700 nRad as in~\cite{CQGA}) and $\theta_1$ the (much smaller) jitter.
In order to obtain a first order expansion in $k a \theta_1$, we need the derivative $G'$ (with respect to $\Omega$) of $G=N/D$ for 
$\Omega_0 \sim 0.62$. 
Owing to the well known properties of Bessel functions, we have
$$
N'_{2n,2m}(\Omega_0,v) \equiv \frac{\partial N_{2n,2m}}{\partial \Omega}(\Omega_0,v) \ = \ - \, \int_0^1 J_{2m+1}(\Omega_0 \rho)R_{2n}^{(2m)}(\rho)
e^{-v \rho^2} \rho^2\,d\rho  \ 
$$ $$
+ \ \frac{2m}{\Omega_0}  \int_0^1 J_{2m}(\Omega_0 \rho)Z_{2n}^{(2m)}(\rho)e^{-v \rho^2}\rho\,d\rho
$$
and
$$
D'(\Omega_0,v) \equiv \frac{\partial D}{\partial \Omega}(\Omega_0,v)  \ = \
- \ \int_0^1 J_1(\Omega_0) e^{-v \rho^2}\rho^2\,d\rho ~,
$$
so that the coefficient we need to evaluate the spectral density of the noise due to
the jitter is determined (for Zernike indices $(2n,2m)$) by:
$$
\frac{\partial}{\partial \Omega} \delta \Phi_{2n,2m} (\Omega_0) \ = \ \left[ \frac{\sigma_{2n,2m}}{\lambda} \cos 2m\psi
+ \frac{\sigma'_{2n,2m}}{\lambda} \sin2m\psi \right] \times 2\pi  c_{2n}^{(2m)} ka G'_{n,m}\theta_1
$$
with (see Eq.~\ref{defr}) $G'_{n,m}\ = \ (N'_{2n,2m}D-N_{2n,2m}D')/D^2$. 
The desired spectral density of the noise $S_{\delta\Phi}^{1/2}$ due to the jitter $\theta_1$ is related to the spectral
density $S_{\theta_1}^{1/2}$ by:
$$
S_{\delta \Phi}^{1/2} (f) \ = \  \sum_{n,m} \left[ \frac{\sigma_{2n,2m}}{\lambda} \cos2m\psi+
 \frac{\sigma'_{2n,2m}}{\lambda}\sin2m\psi\right]
k\, \gamma_{2n}^{(2m)}(\Omega_0) \times S_{ \theta_1}^{1/2}(f)
$$
with the scaling lengths:
$$
\gamma_{2n}^{(2m)}(\Omega_0) \equiv  2\pi  c_{2n}^{(2m)} aG'_{n,m} ~.
$$
If now we need the spectral density of the equivalent displacement $\delta L$, we simply have 
$$
S_{\delta  L}^{1/2} (f) \ = \  \frac{\lambda}{2\pi} S_{\delta \Phi}^{1/2} (f) \ = \ 
$$ $$ \sum_{n,m}
\left[\frac{\sigma_{2n,2m}}{\lambda}\cos2m\psi+  \frac{\sigma'_{2n,2m}}{\lambda}\sin2m\psi \right]  \gamma_{2n}^{(2m)}(\Omega_0) \times S_{ \theta_1}^{1/2}(f) ~ .
$$
Table \ref{tab1} gives the coefficients $\gamma_{2n}^{(2m)}(\Omega_0)$ for the first Noll indices (units are [m]). Noll indices are
frequently employed to have a one-index list of Zernike functions~\cite{nollind}. We give the corresponding
$(n,m)$ Zernike indices. The coefficients are computed for the two reference values of $a/w$: 1.12 and 1.50.

\begin{figure}[t]
\centering
\centerline{\includegraphics[width=16cm]{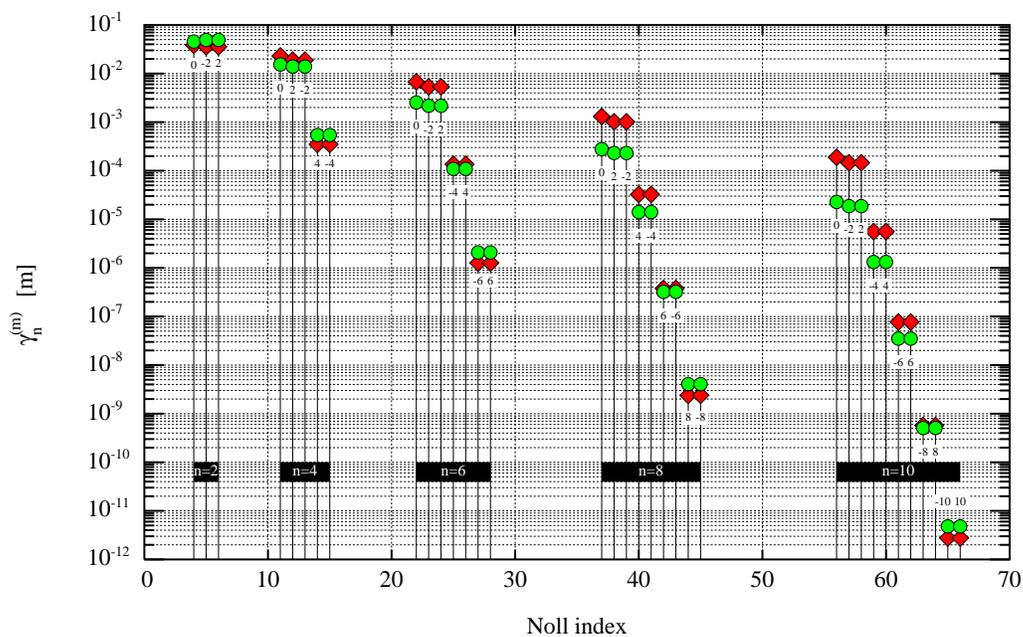}}
\caption{Values of $\gamma_n^{(m)}(\Omega_0)$ [m], for $a/w=$1.5 (red diamonds, and $a/w=$1.12
(green circles) versus Noll index $N$~\cite{nollind} and corresponding $(n,-n \le m \le n)$ Zernike indices.}
\label{fig:scaling_plot}
\end{figure}

\begin{figure}[t]
\centerline{\includegraphics[width=16cm]{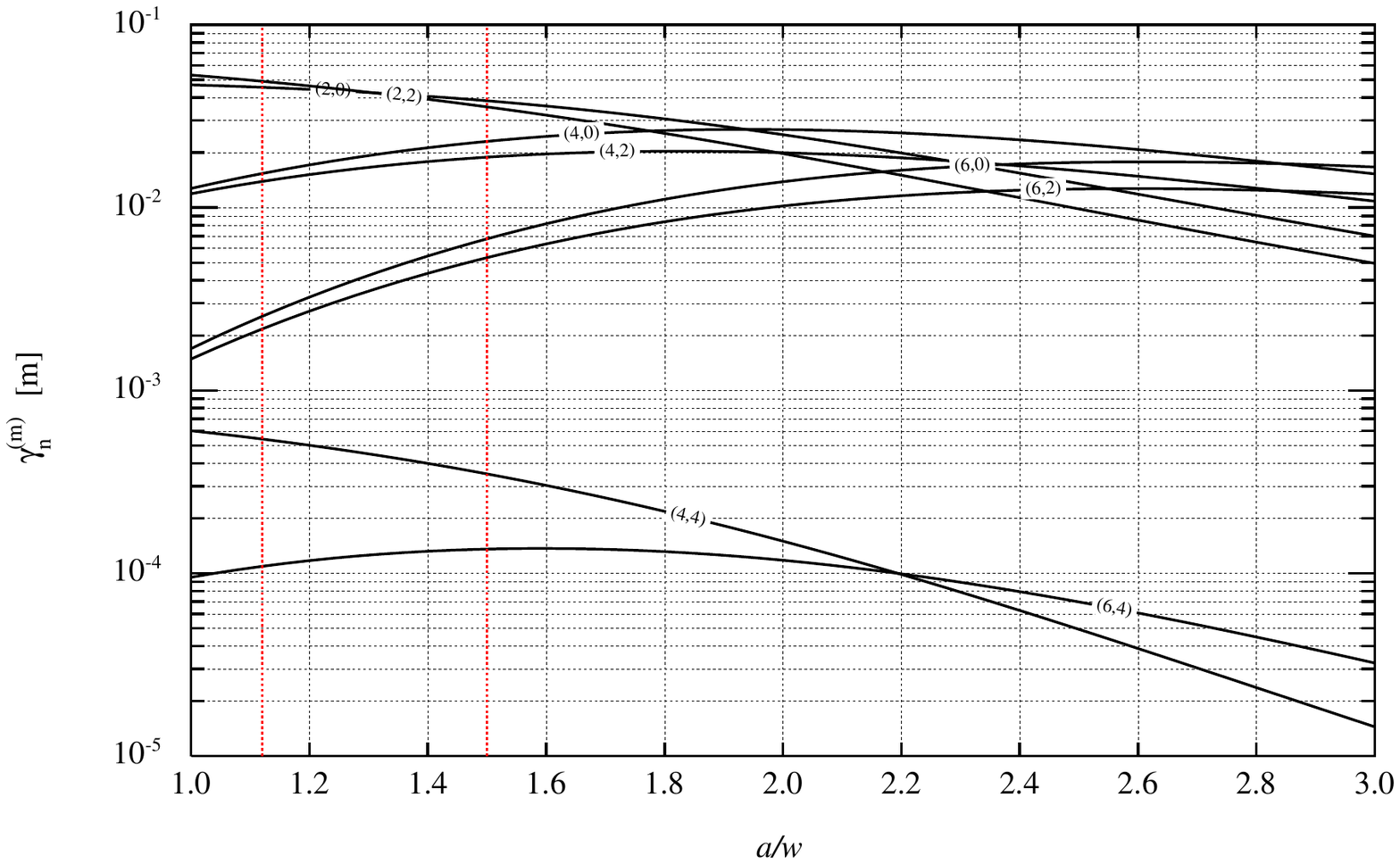}}
\caption{Dependence of $\gamma_n^{(m)}$ on the ratio $a/w$ (telescope aperture/beam width).}
\label{fig:gamma}
\end{figure}

\begin{table}\label{tab1}
\center
\begin{tabular}{| l | l | l | l |}
 \hline \hline
$N : $  &  $(n,|m|)$  &  $\gamma_n^{(m)}$[m]  &  $\gamma_n^{(m)}[m] $ \\ 
      &   & $a/w$=1.12 &  $a/w$ =1.5  \\ \hline
4 & ( 2, 0)  & 0.4553E-01 & 0.3843E-01   \\
6 & ( 2, 2) &  0.4913E-01 & 0.3558E-01 \\ \hline
11 & ( 4, 0) & 0.1537E-01 & 0.2304E-01 \\
12 & ( 4, 2) & 0.1391E-01 & 0.1889E-01 \\
14 & ( 4, 4) & 0.5422E-03  & 0.3494E-03\\ \hline
22  & ( 6, 0) & 0.2542E-02  & 0.6750E-02 \\
 24 & ( 6, 2) & 0.2169E-02  & 0.5327E-02 \\
 26 &  ( 6, 4) & 0.1091E-03 &  0.1356E-03 \\
 28 & ( 6, 6) & 0.2078E-05  & 0.1266E-05 \\ \hline
 37 &( 8, 0) & 0.2783E-03  & 0.1307E-02 \\
 38 & ( 8, 2) & 0.2313E-03  & 0.1015E-02 \\
 40 & ( 8, 4) &  0.1408E-04  & 0.3251E-04 \\
 42 & ( 8, 6) & 0.3214E-06 & 0.3772E-06 \\
 44 & ( 8, 8) & 0.4053E-08  & 0.2389E-08 \\ \hline
 56 & (10, 0) & 0.2276E-04  & 0.1891E-03 \\
 58 & (10, 2) & 0.1869E-04 &  0.1458E-03 \\
 60 & (10, 4) & 0.1324E-05  & 0.5589E-05 \\
 62 & (10, 6) & 0.3504E-07   & 0.7695E-07 \\
 64 & (10, 8) & 0.5068E-09  & 0.5723E-09 \\
 66 & (10,10) & 0.4785E-11 &  0.2762E-11 \\ 
\hline \hline
\end{tabular}
\caption{Scaling lengths $\gamma_n^{(m)}$ (units = [m]) for Noll indices $N$~\cite{nollind} equivalent to
Zernike indices $(n,m)$ for two different $a/w$ ratios.}
\end{table}
We also show in Fig.~\ref{fig:scaling_plot} the same data plotted on a logarithmic scale.
Fig.~\ref{fig:gamma} gives an idea of the dependence of the most significant scale factors $\gamma_{2n}^{(2m)}$ with respect to the
ratio $a/w$.
In the very worst case, if all aberrations accumulate with the same magnitude $\sigma$ and identical signs, 
and assuming $\psi=0$, the resulting
global scaling factor would be:
$$
\gamma \ \equiv \ \sum_{n,m\ge0} \gamma_{2n}^{(2m)} ~.
$$
The sum is formally infinite, but converges in practice because of the rapid decrease of the $\gamma_{2n}^{(2m)}$.
For example the result obtained for Noll indices $ N \le 28$ and
with a ratio $a/w = 1.12$ is $\gamma \sim 0.13$m; with a ratio $a/w=1.5$ the result is similar.

With a noise spectral density of angular jitter $S_{\theta_1}^{1/2}\sim10\ {\rm nRd/\sqrt{Hz}}$, the spectral density of wavefront displacement would be roughly:
\begin{equation}\label{last}
S_{\delta  L}^{1/2} (f) \ = \ \frac{\sigma}{\lambda}\times \gamma \times 10^{-8} {\rm m/\sqrt{Hz}}
\ \simeq \  \frac{\sigma}{\lambda} \, \times 1300 \  {\rm pm/\sqrt{Hz}} ~ ,
\end{equation}
slightly larger than the result given in~\cite{CQGA}, Eq. 48, where the factor was 1200 ${\rm m/\sqrt{Hz}}$ instead of 1300 ${\rm m/\sqrt{Hz}}$. See in the following Annex, Sec.~\ref{sec:annex}, a detailed comparison between the present numerical results and the estimations given in our 
preceding paper~\cite{CQGA}. 

We can also compare our order of magnitude calculations with the estimates given by Sasso et al.~\cite{sasso_2018}. Their estimate, given by their Table 3, is $\delta L$ = 0.07 pm/nrad for an amplitude of Zernike of $\lambda/20$. If we look at our
Eq.~\ref{last}, and ignoring the spectral densities, with
$\sigma/\lambda \sim$ 1/20, one has $\delta L \sim$ 65 pm for $\theta_1$ = 10 nrad, or 6.5 pm/nrad, which
is two orders of magnitude higher. 
Also, we have used 700 nrad for the static mispointing where a value of 10 nrad is used in ~\cite{sasso_2018}.
However we must keep in mind that we have cumulated all aberrations in a worst case scenario,
whereas their approach is based on Monte-Carlo methods, in which
some defects can randomly cancel others.
%===================================

\section{Conclusion}
\label{sec:conclusion}
It appears that the estimations based on an expansion in $\Omega_0$ limited to the second order, as presented in our preceding paper~\cite{CQGA}, slightly underestimate
the noise by a few percent for the first Zernike contributions. The global result obtained for worst case conditions is
also slightly higher, due to contributions of several (small) higher order terms not taken into account in our preceding theoretical presentation.
On the other hand, our preceding paper gave no estimations of scaling factors values for $m\ge4$, whereas the
numerical treatment presented here does, showing that those values are much smaller for higher order Zernike $n  > 2$, as expected.

The attitude jitter of a SC can induce a noise in the distance determination of the associated laser link. This is not the only example of tilt-to-lengh coupling, which can also arise in the interferometeric system for a LISA test mass because of SC jitter with respect to the reflected laser beam from the test mass. In addition, noise can come from the jitter at the local SC with respect to the laser beam from a far SC~\cite{troebs}.
Consequently, wavefront error, as discussed in this article, contributes only to a part of the total tilt-to-length coupling for LISA.

The subject of wavefront errors and beam jitter in the LISA optical system has, of course, been the subject of other studies~\cite{sasso_2018,sasso_2018b}. For many years this has been recognized as an important problem for LISA~\cite{bender_2015}.

As discussed in this paper, the analysis and study of the far-field laser phase and intensity are critical for ensuring that the LISA GW detector operates at its desired sensitivity. The numerical results presented in this paper extend our previous analytic work~\cite{CQGA}, and help to display how aberrations in the LISA telescope plus pointing errors can create phase noise after the beam has traveled 2.5 Mkm.
%===============================================================
\section{Annex}
\label{sec:annex}
In order to compare the numerical results presented here with the analytic results of our preceding paper~\cite{CQGA}, simply take the very first Noll indices~\cite{nollind} and a ratio $a/w=1.5$.
For $n=2,m=0$, and with a noise spectral density of angular jitter of 10nRad/$\rm \sqrt{Hz}$, we find
$$
S_{\delta  L,(2,0)}^{1/2} (f)\ =\ \frac{\sigma_{2,0}}{\lambda} \times  3.843\times 10^{-2} \times 10^{-8} \, {\rm m/\sqrt{Hz} }\ =
\ \frac{\sigma_{2,0}}{\lambda}\times  3.843 \times 10^{-10} {\rm m}/\sqrt{Hz} ~ .
$$
In our preceding paper~\cite{CQGA}, we had (Eq.47) with a coefficient $\alpha_1=2.425$:
$$
S_{\delta  L,(2,0)}^{1/2} (f)\ = \ \frac{\sigma_{2,0}}{\lambda} \times1.55\times 10^{-10} \times 2.425\,  {\rm m/\sqrt{Hz} } \  = \ 
\frac{\sigma_{2,0}}{\lambda} \times 3.759\times 10^{-10} {\rm m}/\sqrt{Hz} ~ .
$$
For $n=2,m=2$, the same way:
$$
S_{\delta  L,(2,2)}^{1/2} (f)\ = \ \frac{\sigma_{2,2}}{\lambda} \times3.558 \times 10^{-2} \times 10^{-8} \, {\rm m/\sqrt{Hz}} \ =
\  \frac{\sigma_{2,2}}{\lambda} \times3.558 \times 10^{-10} {\rm m}/\sqrt{Hz} ~ ,
$$
in our preceding paper, with a coefficient  $\beta_1=2.247$:
$$
S_{\delta  L,(2,2)}^{1/2} (f)\ = \ \frac{\sigma_{2,2}}{\lambda} \times1.55\times 10^{-10} \times 2.247\,  {\rm m/\sqrt{Hz}}  \  = \ 
\frac{\sigma_{2,2}}{\lambda} \times 3.483\times 10^{-10} {\rm m}/\sqrt{Hz} ~ .
$$
It can be seen that the results are very similar, up to a few percent. This small difference is due to a better (numerical) evaluation of integrals involving Bessel functions. 
%=======================================================
%=======================================================
\section*{Acknowledgements}
The ARTEMIS Laboratory gratefully acknowledges the support of the Centre National d'\'{E}tudes Spatiales (CNES).
%=======================================================

%\bibliographystyle{ws-rv-van}
%\bibliography{ws-rv-sample}
                        
\end{document}